\documentclass[aip,reprint]{revtex4-1}
\usepackage{graphicx,amsmath,epstopdf, natbib}
\draft 
\usepackage[bookmarks=false]{hyperref}
\begin{document}

\title{Fast, low-power manipulation of spin ensembles in superconducting microresonators} 

\author{A. J. Sigillito}
\email[]{asigilli@princeton.edu}
\affiliation{Department of Electrical Engineering, Princeton University, Princeton, New Jersey 08544, USA}

\author{H. Malissa}
\altaffiliation{Current address: Department of Physics and Astronomy, University of Utah, Salt Lake City, Utah 84112, USA}
\affiliation{Department of Electrical Engineering, Princeton University, Princeton, New Jersey 08544, USA}

\author{A. M. Tyryshkin}
\affiliation{Department of Electrical Engineering, Princeton University, Princeton, New Jersey 08544, USA}

\author{H. Riemann, N. V. Abrosimov}
\affiliation{Institut f{\"u}r Kristallz{\"u}chtung, D-12489 Berlin, Germany}

\author{P. Becker}
\affiliation{Physikalisch-Technische Bundesanstalt, D-38116 Braunschweig, Germany}

\author{H.-J. Pohl}
\affiliation{VITCON Projectconsult GMBH, D-07745 Jena, Germany}

\author{M. L. W. Thewalt}
\affiliation{Department of Physics, Simon Fraser University, Burnaby, British Columbia V5A 1S6, Canada}

\author{K. M. Itoh}
\affiliation{School of Fundamental Science and Technology, Keio University, Yokohama, Kanagawa 2238522, Japan}

\author{J. J. L. Morton}
\affiliation{London Centre for Nanotechnology, University College London, London WC1H 0AH, UK}

\author{A. A. Houck}  
\affiliation{Department of Electrical Engineering, Princeton University, Princeton, New Jersey 08544, USA}

\author{D. I. Schuster}
\affiliation{Department of Physics and James Franck Institute, University of Chicago, Chicago, Illinois 60637, USA}

\author{S. A. Lyon}
\affiliation{Department of Electrical Engineering, Princeton University, Princeton, New Jersey 08544, USA}


\date{\today}

\begin{abstract}
We demonstrate the use of high-Q superconducting coplanar waveguide (CPW) microresonators to perform rapid manipulations on a randomly distributed spin ensemble using very low microwave power (400 nW). This power is compatible with dilution refrigerators, making microwave manipulation of spin ensembles feasible for quantum computing applications. We also describe the use of adiabatic microwave pulses to overcome microwave magnetic field ($B_{1}$) inhomogeneities inherent to CPW resonators. This allows for uniform control over a randomly distributed spin ensemble. Sensitivity data are reported showing a single shot (no signal averaging) sensitivity to $10^{7}$ spins or $3 \times 10^{4}$ spins/$\sqrt{Hz}$ with averaging.
\end{abstract}

\pacs{}

\maketitle 


Superconducting coplanar waveguide (CPW) resonators are a good alternative to conventional volume resonators for many applications because of their high sensitivity, low power requirements, and small size\cite{cite09,cite06,cite05}. They are of particular interest in the construction of hybrid quantum systems utilizing the long coherence times of spin-based qubits and the strong coupling of superconducting qubits\cite{cite07,cite11,cite12,cite10}. Hybrid quantum systems have been realized for nitrogen-vacancy (N-V) centers coupled to transmon qubits \cite{cite07}. These systems are limited by dephasing of the spin ensemble ($T_{2}^{*}$) which is often hundreds of nanoseconds in solids. This can be improved by employing refocusing techniques common in pulsed electron spin resonance (ESR), enabling spin memories effective over the full coherence time of the electron spin ($T_{2}$), which can be 10 s in Si\cite{tyryshkin2012}. However, refocusing pulses have been difficult to implement because the microwave magnetic field ($B_{1}$) in a CPW is inhomogeneous. Furthermore, it has been suggested that driving ensemble rotations with microwave pulses is too slow for quantum computing and can lead to excessive microwave heating of the system\cite{cite03}. In this letter we address these issues. We report CPW resonators capable of performing $\pi$-rotations on a spin ensemble in 40 ns while using powers compatible with dilution refrigerators (40 $\mu W$ peak). We also present data showing the use of adiabatic microwave pulses to overcome $B_{1}$ inhomogeneities, enabling accurate spin manipulations over a randomly distributed ensemble.

The spin ensembles discussed in this letter are donor electron spins in Si. Donors in Si have the advantage over N-Vs of long coherence times exceeding seconds\cite{tyryshkin2012, wolfowicz2013} and a wealth of experience in fabrication techniques. However, only donors with large zero-field splittings are compatible with these hybrid systems because superconducting qubits have a low critical field. Bismuth donors in Si have been proposed as a good candidate for coupling to superconducting qubits due to their long coherence times, existence of clock transitions, and large zero-field splitting\cite{george2010}. While the data presented in this letter focuses on P donors, the methodology and results are not unique to P and easily extended to other spin ensembles.

One major challenge to performing rotations on a spin ensemble at low temperature is microwave heating. A typical X-band pulsed spectrometer performing 40 ns $\pi$-rotations can require an input power of tens of watts for a high quality factor (Q) volume resonator or up to a kilowatt of power for the lower-Q resonators used in studying systems with short coherence times. Most of that energy is reflected from the resonator, but heating can be significant. However, the resonators we report have a substantially smaller mode volume, so less power is required to drive ensemble rotations. The small mode volume is possible because CPW resonators require only one dimension to be on the order of the resonant frequency wavelength. The other two dimensions can be made arbitrarily small.

Another challenge to manipulating an ensemble of randomly distributed spins is $B_{1}$ inhomogeneity, which is intrinsic to CPW designs. Field inhomogeneities lead to non-uniform control over a macroscopic ensemble, because the tipping angle of a given spin is proportional to the driving field strength.  Spins in regions of large $B_{1}$ will be rotated more than spins in regions of small $B_{1}$. There are essentially two approaches to overcoming these inhomogeneities. The first is to tailor the device geometry such that spins are located in regions of homogeneous $B_{1}$. This is accomplished by either changing the resonator structure \cite{cite01} or by confining the ensemble to a small region where $B_{1}$ is uniform. These methods typically require that the volume of the spin ensemble be smaller than the mode volume of the resonator, leading to weaker coupling. The second approach is to construct microwave pulses that compensate for $B_{1}$ inhomogeneities. This allows for uniform control over an ensemble filling nearly the entire mode volume of the resonator. We have chosen the latter method and utilize adiabatic microwave pulses, which produce $B_{1}$-insensitive spin rotations. Such adiabatic pulses are known in the nuclear magnetic resonance community, and we have tested several varieties\cite{cite02, cite04, cite08}. The best results were obtained by combining a WURST-20 (Wideband, Uniform Rate, Smooth Truncation 20) envelope shape\cite{cite08} with a BIR-4 ($B_{1}$ Insensitive Rotation 4) phase compensation\cite{cite04}. The WURST-20 envelope shapes the pulse as $sin^{20}(\pi t /t_{p})$ where $t$ is time and $t_{p}$ is the pulse length. The BIR-4 technique breaks the WURST-20 pulse in half and combines four of these waveforms in a time-reversed order. The BIR-4 technique is robust against off-resonance effects, and specifically compensate for geometrical phase errors. Individually, BIR-4 and WURST-20 have been discussed in the literature\cite{cite02,cite04,cite08}. 

\begin{figure}
\includegraphics{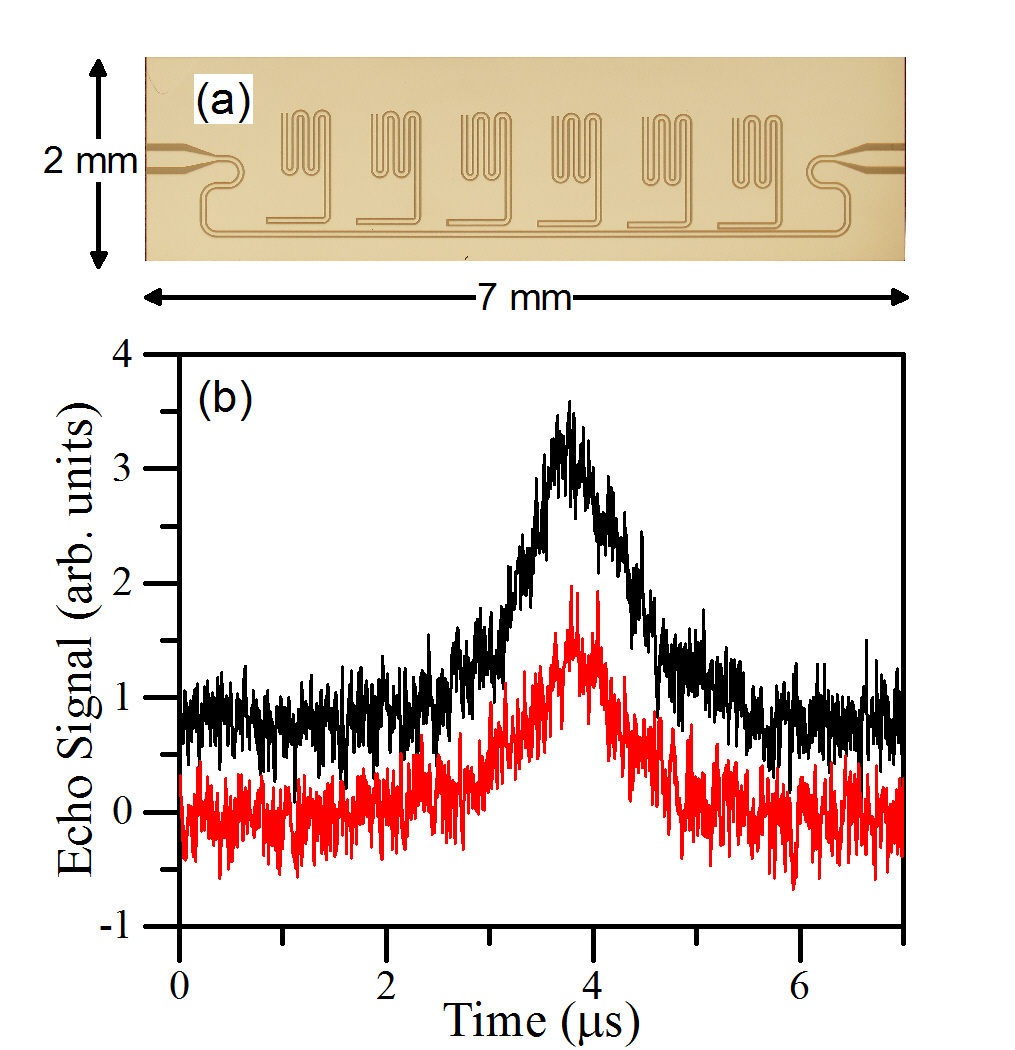}
\caption{\label{fig:fig1} (a) Optical micrograph of the device. Six resonators (serpentine structures) are capacitively coupled to a common transmission line. (b) Single shot spin echoes acquired using adiabatic (top black) and rectangular (bottom red) pulses. The adiabatic pulse echo has been shifted by 0.8 for clarity. Data were taken at 1.7 K in a DC magnetic field of 0.26 T.} 
\end{figure}

The spin ensemble consisted of a 25 $\mu$m epitaxial layer of $^{28}$Si grown on high resistivity p-type Si. The epi-layer was doped to a concentration of $8 \times 10^{14}$ P donors/cm$^{3}$. This layer had 50 nm of Al$_{2}$O$_{3}$ grown on the surface to protect against a SF$_{6}$ plasma used in the device fabrication. Six CPW resonators were patterned in 50 nm thick Nb films directly on the Al$_{2}$O$_{3}$ surface and they are shown in Fig.~\ref{fig:fig1}a. The fabrication techniques have been described previously\cite{cite09}. The CPW center conductor width was 30 $\mu$m, with a gap width of 17.4 $\mu$m defining an impedance of 50 $\Omega$. Each resonator had a unique frequency, spanning a range from 7 GHz - 8 GHz, and all were nearly critically coupled to a common transmission line. Most of the results reported in this letter were obtained using one resonator with a 7.17 GHz center frequency, Q of $\sim$2000, and coupling coefficient of 1.15. Resonators were wire bonded to copper printed circuit boards equipped with microwave connectors and cooled to 1.7 K. The output of the resonator transmission line was attached to a low-noise cryogenic preamplifier (Caltech LNA 1-12). We applied a direct current (DC) magnetic field to the sample, taking care that the plane of the Nb film remained parallel to the field. Careful alignment prevented the trapping of magnetic flux vortices in the superconducting film, which are lossy and serve as a decoherence mechanism\cite{cite09}. 

Two-pulse Hahn echo experiments (nominally $\pi$/2(+x) -- $\tau$ -- $\pi$(+y) -- $\tau$ -- echo)  were conducted with a delay time ($\tau$) of 15 $\mu$s and the results are shown in Fig. ~\ref{fig:fig1}b. Because the relaxation time (T$_{1}$) of P donors at 1.7 K is on the order of minutes, the back side of the sample was illuminated with a 1050 nm light emitting diode for 50 ms prior to each two pulse experiment. The light relaxes the spins allowing fast repetition rates. For 400 ns rectangular $\pi$-pulses, the optimal microwave power was $-34$ dBm (400 nW). The experiment was repeated using BIR-4-WURST-20 adiabatic pulses. For our devices, the optimal adiabatic pulse chirped from 2 MHz below to 2 MHz above the resonant frequency in 11 $\mu$s. This chirp bandwidth was chosen to be wider than the ESR linewidth of 0.3 MHz in order to excite the entire spin ensemble. The corresponding peak microwave power was $-30$ dBm (1 $\mu$W). The integrated signal-to-noise (S/N) ratio for the single shot (no signal averaging) rectangular and adiabatic pulse experiments are 84 and 146, respectively. Thus, by using adiabatic pulses the signal increased by a factor of 1.74.

To demonstrate that microwave manipulation of spin ensembles can occur on timescales compatible with quantum computing, shorter rectangular pulses were tested on a second device. The device had a Q of 3200, center frequency of 7.14 GHz, and coupling coefficient of 1.15. The optimal power for 400 ns and 40 ns rectangular $\pi$-pulses was $-33$ dBm and $-13$ dBm, respectively. We expect the 40 ns $\pi$-pulse to be distorted by the high-Q resonator. However, a 20 dB increase in power still led to order of magnitude shorter excitation pulses.

\begin{figure}
\includegraphics{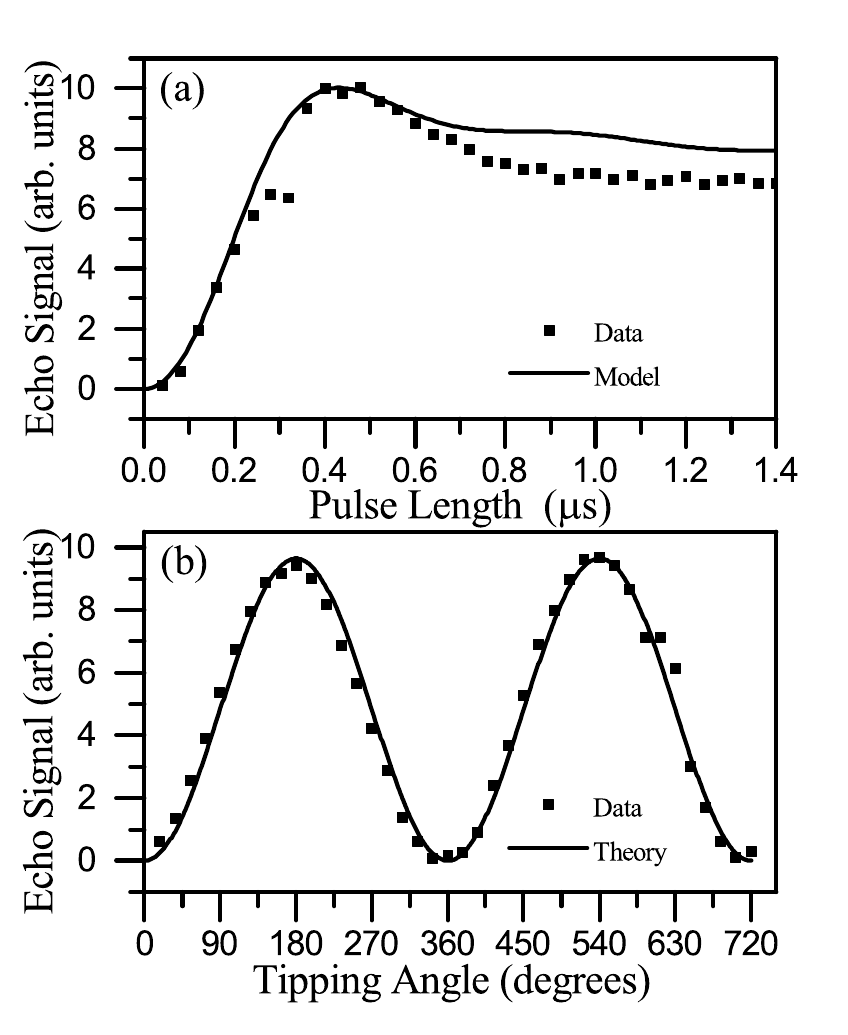}
\caption{\label{fig:fig2}Echo intensity as a function of tipping angle for (a) rectangular pulses and (b) adiabatic pulses. Experimental data is represented by solid squares. The curve in (a) is the best fit curve from the model (Eq. 2) and the curve in (b) assumes ideal spin rotations. Data were taken at 1.7 K in a DC magnetic field of 0.26 T.} 
\end{figure}

Two-pulse experiments (nominally $\pi$/2(+x) -- $\tau$ -- $\theta$(+y) -- $\tau$ -- echo) were also performed where the tipping angle of the second pulse was varied. For rectangular pulses, a microwave power of $-27$ dBm was used, with a 200 ns first pulse. The second pulse was varied from 0 ns to 1400 ns. The echo intensity as a function of pulse length is plotted in Fig.~\ref{fig:fig2}a. This is compared to a similar experiment conducted using adiabatic pulses, shown in Fig.~\ref{fig:fig2}b. When using adiabatic pulses, tipping angles were well defined such that the first pulse performed a $\pi$/2 rotation and the second pulse tipping angle varied from 0 to 4$\pi$. An optimal peak microwave power of $-30$ dBm was used for these experiments. It is clear from the data that the $B_{1}$ inhomogeneity greatly affects the rectangular pulse experiment, which shows no Rabi oscillations, while the adiabatic pulses produce Rabi oscillations as expected. 

To understand these experiments, a model was developed to simulate the results. The normalized $B_{1}$ distribution in the CPW resonator was computed using a conformal mapping technique, and the echo intensity was determined by summing over the contribution of each individual spin, as previously described\cite{cite09}.  The contribution of a single spin to the echo is given by \begin{equation}
signal(\boldsymbol{r}) = g_{s}(\boldsymbol{r}) sin(\theta_{1}(\boldsymbol{r})) sin^{2}(\theta_{2}(\boldsymbol{r})/2)
\end{equation} where $g_{s}(\boldsymbol{r})$ is the coupling of a spin at position $\boldsymbol{r}$ to the resonator, $\theta_{1}(\boldsymbol{r})$ is the tipping angle of the first pulse (the first pulse is nominally $\pi$/2, but the actual tipping angle varies with spin position), and $\theta_{2}(\boldsymbol{r})$ is the tipping angle of the second pulse. For rectangular pulses, the tipping angle is $g\mu_{B} B_{1}(\boldsymbol{r}) t_{p} / \hbar$, where $g$ is the electron g-factor, $\mu_{B}$ is the Bohr magneton, and $\hbar$ is the reduced Planck constant. The spin-resonator coupling is linearly proportional to $B_{1}(\boldsymbol{r})$, the microwave magnetic field. We can write $B_{1}(\boldsymbol{r}) = C B_{1n}(\boldsymbol{r})$ where $B_{1n}$ is the normalized $B_{1}$, and $C$ depends on the microwave power, cavity coupling, and Q. Thus, by writing $g_{s}(\boldsymbol{r}) = A B_{1n}(\boldsymbol{r})$, the total signal becomes

\begin{multline} 
signal = \int d\boldsymbol{r} A C B_{1n}(\boldsymbol{r})sin\biggl(\frac{g \mu C B_{1n}(\boldsymbol{r}) t_{1}}{\hbar}\biggr) \\ sin^{2}\biggl(\frac{g \mu C B_{1n}(\boldsymbol{r}) t_{2}}{2 \hbar}\biggr)
\end{multline} where $t_{1}$ is the first pulse duration and $t_{2}$ is the second pulse duration. The constant, $A$, simply normalizes the vertical scale whereas $C$ determines the shape of the curve shown in Fig.~\ref{fig:fig2}a. By varying $C$ for a given microwave power, we obtain a good fit to the data. From $C$, we compute $B_{1}(\boldsymbol{r})$, which includes resonator Q, losses, and coupling.

\begin{figure}
\includegraphics{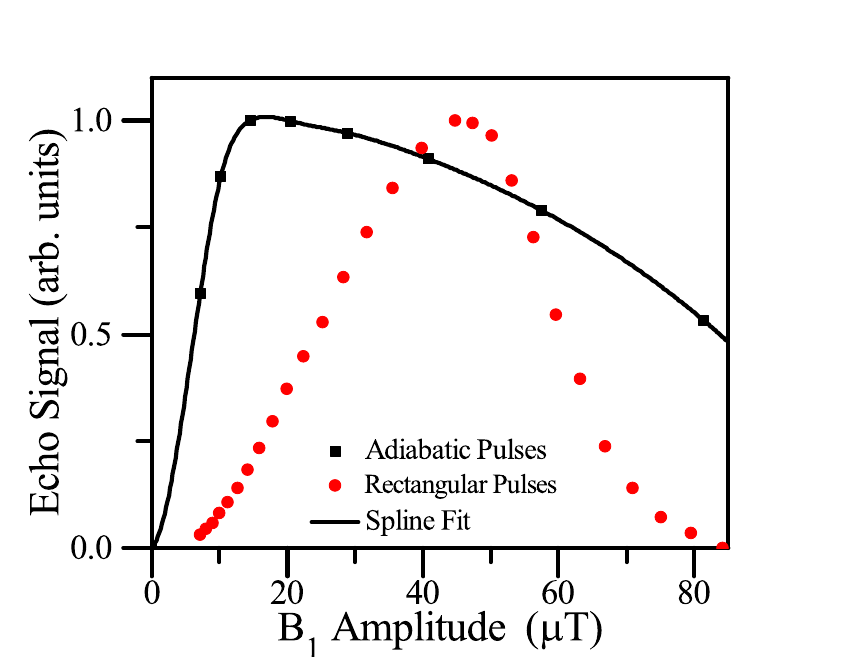}
\caption{\label{fig:fig3}Plot of echo amplitude as a function of B$_{1}$ for a sample in a high-homogeneity resonator with adiabatic pulses (black) and rectangular pulses (red). Data were taken at 4.6 K in a DC magnetic field of 0.34 T. The solid line is a spline fit to the data and is used later in a simulation. } 
\end{figure}

To evaluate the performance of adiabatic pulses quantitatively, we performed echo experiments using a high-homogeneity commercial dielectric resonator (Bruker MD5). We used a bulk doped $^{28}$Si crystal\cite{cite14} with $3.3 \times10^{15}$ P donors/cm$^{3}$. The sample volume was 1 mm $\times$ 2 mm $\times$ 4 mm, and $B_{1}$ homogeneity varied by no more than 5\% over the sample volume.  In these experiments, adiabatic pulse tipping angles were defined as $\pi$/2 and $\pi$, while the microwave power and thus $B_{1}$ was varied. The integrated echo intensity as a function of $B_{1}$ is shown in Fig.~\ref{fig:fig3}. As a comparison, the experiment was repeated using rectangular pulses with a $\pi$-pulse width of 400 ns ($B_{1}\sim$10 times the ESR linewidth), and these data are also shown.  At least half of the maximum echo intensity is observed for $B_{1}$ in the range of 6 $\mu$T to 83 $\mu$T for adiabatic pulses, while the range is 24 $\mu$T to 61 $\mu$T for rectangular pulses. This comparison shows that adiabatic pulses correct $B_{1}$ inhomogeneity over an order of magnitude (two orders of magnitude in microwave power).

\begin{figure}
\includegraphics{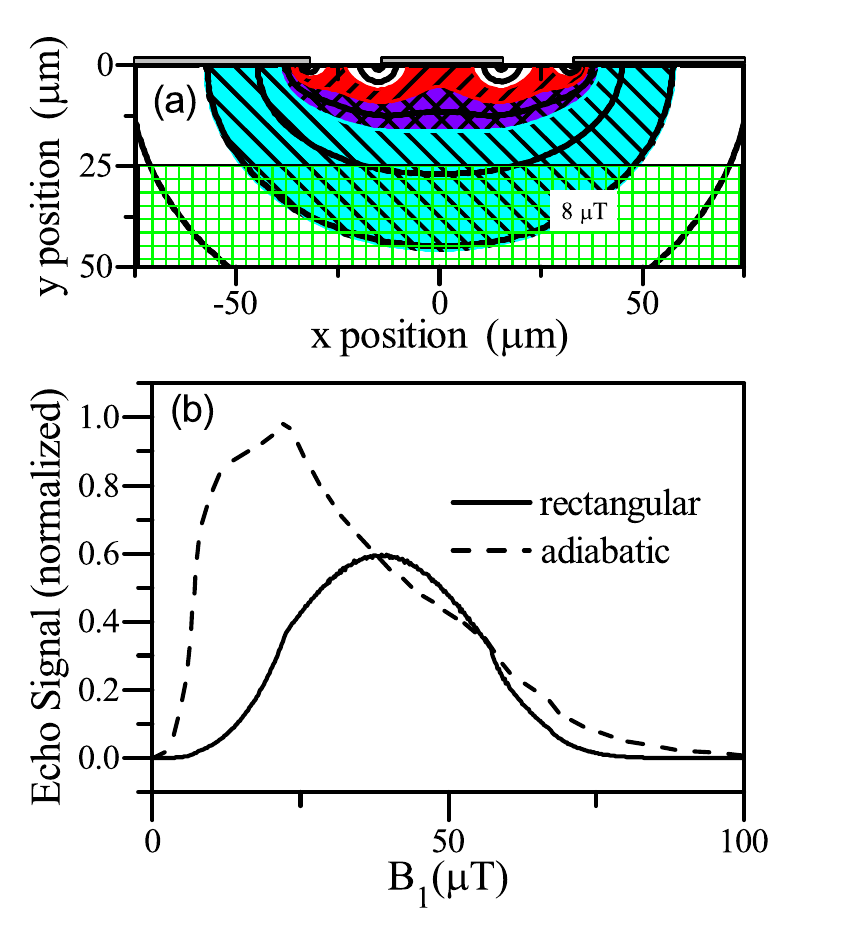}
\caption{\label{fig:fig4}(a) Cross section of resonator at an antinode in $B_{1}$ with contour lines indicating $B_{1}$ magnitude for a 400 nW input power. The hatched regions denote the location of spins contributing to 2/3 of the echo intensity for rectangular (red) pulses and adiabatic (blue) pulses (violet where they overlap). The green-hatched region below 25 $\mu$m denotes the undoped portion of the sample. (b) Plot of the echo intensity contribution of all spins at particular values of $B_{1}$.} 
\end{figure}

Combining the simulated $B_{1}$ distribution with our measurements of echo intensity as a function of $B_{1}$ (Fig.~\ref{fig:fig3}), we identified the regions of the sample contributing most to the echo signal. Fig.~\ref{fig:fig4}a is a cross section of the CPW resonator at an antinode in the magnetic field (near the shorted end of the resonator). The CPW is depicted at the top of the plot, and the magnitude of $B_{1}$ in the sample is shown with contours (the 8 $\mu$T contour is labeled, and $B_{1}$ for each subsequent contour increases by a factor of two). The hatched regions in the figure denote where 2/3 of the signal originates for the rectangular and adiabatic pulses. The Si sample used for these experiments had a 25 $\mu$m thick P-doped $^{28}$Si epi-layer, and thus the ensemble volume only extends down to the green cross-hatched region. Using the $B_{1}$ distribution, the contribution of all spins, at each value of $B_{1}$, to the echo was computed and is plotted in Fig.~\ref{fig:fig4}b for both adiabatic and rectangular pulses.  By integrating over these curves we obtained the total signal intensity and found that adiabatic pulses produce a signal that is 1.73 times larger than the signal produced by rectangular pulses. This is in excellent agreement with the value of 1.74 observed in experiment. Note that because adiabatic pulses are sensitive to a volume larger than the epi-layer, the adiabatic pulse signal would increase when using a bulk doped sample. We also note that rectangular pulses are sensitive to a thin region of spins which could allow for high resolution tomography experiments.

From these simulations, we estimate the sample volume coupled to our resonators to be $3.9 \times 10^{-6}$ cm$^{3}$. The doping density of the sample is $8 \times 10^{14}$ donors/cm$^{3}$, so there are $1.6 \times 10^{9}$ spins coupled to the resonator per hyperfine line. We measured these spins in a single shot (using no signal averaging and utilizing a single two-pulse sequence) and found a S/N of 84 using rectangular pulses and 146 with adiabatic pulses. Scaling to S/N = 1, we have sensitivity to $2 \times 10^{7}$ spins when using rectangular pulses and $1 \times 10^{7}$ spins using adiabatic pulses. ESR sensitivity is often reported in units of spins/$\sqrt{Hz}$. These units are appropriate for continuous wave experiments. However, in pulsed electron spin resonance, this sensitivity is limited by the shot repetition rate. Our typical shot repetition rate is 100 Hz (determined by optical spin relaxation) giving a sensitivity of $10^{6}$ spins/$\sqrt{Hz}$. By employing refocusing pulses in a CPMG (Carr-Purcell-Meiboom-Gill) sequence, much faster repetition rates have been achieved\cite{cite13}. Our projected sensitivity when using rectangular pulses in a CPMG sequence is $3 \times 10^{4}$ spins/$\sqrt{Hz}$. This represents an order of magnitude improvement over recently reported values for spin resonance detected by a superconducting qubit\cite{cite06} and is on par with sensitivities reported using surface loop-gap microresonators\cite{cite13}. This value should improve by another factor of 5 by measuring at lower temperatures where spins are fully polarized.

In summary, we have demonstrated the use of superconducting CPW resonators to perform pulsed electron spin resonance using an ultra low power of -34 dBm (400 nW) with a $\pi$-pulse length of 400 ns. We also verify that 40 ns $\pi$-rotations can be achieved using peak powers of about 50 $\mu$W, making CPW resonators compatible with dilution refrigerators. We report a single-shot sensitivity to $10^{7}$ spins or $3 \times 10^{4}$ spins/$\sqrt{Hz}$. This is comparable to the best results reported thus far. Finally, we have shown that BIR-4-WURST-20 pulses can be used to compensate for $B_{1}$ inhomogeneities spanning an order of magnitude. These adiabatic pulses improved our S/N by a factor of 1.74 and substantially improve the uniformity of microwave manipulations of a randomly distributed spin ensemble.



%
%

%

\begin{acknowledgments}
Work at Princeton and UCL was supported in part by the NSF and EPSRC through the Materials World Network Program (DMR-1107606 and EP/I035536/1).  Work at Princeton was also supported by the ARO (W911NF-13-1-0179) and Princeton MRSEC (DMR-0819860), at Keio by the Grant-in-Aid for Scientific Research and Project for Developing Innovation Systems by the Ministry of Education, Culture, Sports, Science and Technology, the FIRST Program by the Japan Society for the Promotion of Science, and the Japan Science and Technology Agency/UK EPSRC (EP/H025952/1), at LBNL by the US Department of Energy (DE-AC02-05CH11231) and the NSA (100000080295), at The University of Chicago by DARPA (N66001-11-1-4123) and NSF through the Chicago MRSEC (DMR-0820054), and at Simon Fraser University by the Natural Sciences and Engineering Research Council of Canada. J.J.L.M. is supported by the Royal Society.
\end{acknowledgments}

\bibliography{resonatorreferences}

\providecommand{\noopsort}[1]{}\providecommand{\singleletter}[1]{#1}%
\begin{thebibliography}{17}%
\makeatletter
\providecommand \@ifxundefined [1]{%
 \@ifx{#1\undefined}
}%
\providecommand \@ifnum [1]{%
 \ifnum #1\expandafter \@firstoftwo
 \else \expandafter \@secondoftwo
 \fi
}%
\providecommand \@ifx [1]{%
 \ifx #1\expandafter \@firstoftwo
 \else \expandafter \@secondoftwo
 \fi
}%
\providecommand \natexlab [1]{#1}%
\providecommand \enquote  [1]{``#1''}%
\providecommand \bibnamefont  [1]{#1}%
\providecommand \bibfnamefont [1]{#1}%
\providecommand \citenamefont [1]{#1}%
\providecommand \href@noop [0]{\@secondoftwo}%
\providecommand \href [0]{\begingroup \@sanitize@url \@href}%
\providecommand \@href[1]{\@@startlink{#1}\@@href}%
\providecommand \@@href[1]{\endgroup#1\@@endlink}%
\providecommand \@sanitize@url [0]{\catcode `\\12\catcode `\$12\catcode
  `\&12\catcode `\#12\catcode `\^12\catcode `\_12\catcode `\%12\relax}%
\providecommand \@@startlink[1]{}%
\providecommand \@@endlink[0]{}%
\providecommand \url  [0]{\begingroup\@sanitize@url \@url }%
\providecommand \@url [1]{\endgroup\@href {#1}{\urlprefix }}%
\providecommand \urlprefix  [0]{URL }%
\providecommand \Eprint [0]{\href }%
\providecommand \doibase [0]{http://dx.doi.org/}%
\providecommand \selectlanguage [0]{\@gobble}%
\providecommand \bibinfo  [0]{\@secondoftwo}%
\providecommand \bibfield  [0]{\@secondoftwo}%
\providecommand \translation [1]{[#1]}%
\providecommand \BibitemOpen [0]{}%
\providecommand \bibitemStop [0]{}%
\providecommand \bibitemNoStop [0]{.\EOS\space}%
\providecommand \EOS [0]{\spacefactor3000\relax}%
\providecommand \BibitemShut  [1]{\csname bibitem#1\endcsname}%
\let\auto@bib@innerbib\@empty
\bibitem [{\citenamefont {Malissa}\ \emph {et~al.}(2013)\citenamefont
  {Malissa}, \citenamefont {Schuster}, \citenamefont {Tyryshkin}, \citenamefont
  {Houck},\ and\ \citenamefont {Lyon}}]{cite09}%
  \BibitemOpen
  \bibfield  {author} {\bibinfo {author} {\bibfnamefont {H.}~\bibnamefont
  {Malissa}}, \bibinfo {author} {\bibfnamefont {D.~I.}\ \bibnamefont
  {Schuster}}, \bibinfo {author} {\bibfnamefont {A.~M.}\ \bibnamefont
  {Tyryshkin}}, \bibinfo {author} {\bibfnamefont {A.~A.}\ \bibnamefont
  {Houck}}, \ and\ \bibinfo {author} {\bibfnamefont {S.~A.}\ \bibnamefont
  {Lyon}},\ }\href {\doibase http://dx.doi.org/10.1063/1.4792205} {\bibfield
  {journal} {\bibinfo  {journal} {Review of Scientific Instruments}\ }\textbf
  {\bibinfo {volume} {84}},\ \bibinfo {eid} {025116} (\bibinfo {year}
  {2013})}\BibitemShut {NoStop}%
\bibitem [{\citenamefont {Kubo}\ \emph {et~al.}(2012)\citenamefont {Kubo},
  \citenamefont {Diniz}, \citenamefont {Grezes}, \citenamefont {Umeda},
  \citenamefont {Isoya}, \citenamefont {Sumiya}, \citenamefont {Yamamoto},
  \citenamefont {Abe}, \citenamefont {Onoda}, \citenamefont {Ohshima},
  \citenamefont {Jacques}, \citenamefont {Dr\'eau}, \citenamefont {Roch},
  \citenamefont {Auffeves}, \citenamefont {Vion}, \citenamefont {Esteve},\ and\
  \citenamefont {Bertet}}]{cite06}%
  \BibitemOpen
  \bibfield  {author} {\bibinfo {author} {\bibfnamefont {Y.}~\bibnamefont
  {Kubo}}, \bibinfo {author} {\bibfnamefont {I.}~\bibnamefont {Diniz}},
  \bibinfo {author} {\bibfnamefont {C.}~\bibnamefont {Grezes}}, \bibinfo
  {author} {\bibfnamefont {T.}~\bibnamefont {Umeda}}, \bibinfo {author}
  {\bibfnamefont {J.}~\bibnamefont {Isoya}}, \bibinfo {author} {\bibfnamefont
  {H.}~\bibnamefont {Sumiya}}, \bibinfo {author} {\bibfnamefont
  {T.}~\bibnamefont {Yamamoto}}, \bibinfo {author} {\bibfnamefont
  {H.}~\bibnamefont {Abe}}, \bibinfo {author} {\bibfnamefont {S.}~\bibnamefont
  {Onoda}}, \bibinfo {author} {\bibfnamefont {T.}~\bibnamefont {Ohshima}},
  \bibinfo {author} {\bibfnamefont {V.}~\bibnamefont {Jacques}}, \bibinfo
  {author} {\bibfnamefont {A.}~\bibnamefont {Dr\'eau}}, \bibinfo {author}
  {\bibfnamefont {J.-F.}\ \bibnamefont {Roch}}, \bibinfo {author}
  {\bibfnamefont {A.}~\bibnamefont {Auffeves}}, \bibinfo {author}
  {\bibfnamefont {D.}~\bibnamefont {Vion}}, \bibinfo {author} {\bibfnamefont
  {D.}~\bibnamefont {Esteve}}, \ and\ \bibinfo {author} {\bibfnamefont
  {P.}~\bibnamefont {Bertet}},\ }\href {\doibase 10.1103/PhysRevB.86.064514}
  {\bibfield  {journal} {\bibinfo  {journal} {Phys. Rev. B}\ }\textbf {\bibinfo
  {volume} {86}},\ \bibinfo {pages} {064514} (\bibinfo {year}
  {2012})}\BibitemShut {NoStop}%
\bibitem [{\citenamefont {Huebl}\ \emph {et~al.}(2013)\citenamefont {Huebl},
  \citenamefont {Zollitsch}, \citenamefont {Lotze}, \citenamefont {Hocke},
  \citenamefont {Greifenstein}, \citenamefont {Marx}, \citenamefont {Gross},\
  and\ \citenamefont {Goennenwein}}]{cite05}%
  \BibitemOpen
  \bibfield  {author} {\bibinfo {author} {\bibfnamefont {H.}~\bibnamefont
  {Huebl}}, \bibinfo {author} {\bibfnamefont {C.~W.}\ \bibnamefont
  {Zollitsch}}, \bibinfo {author} {\bibfnamefont {J.}~\bibnamefont {Lotze}},
  \bibinfo {author} {\bibfnamefont {F.}~\bibnamefont {Hocke}}, \bibinfo
  {author} {\bibfnamefont {M.}~\bibnamefont {Greifenstein}}, \bibinfo {author}
  {\bibfnamefont {A.}~\bibnamefont {Marx}}, \bibinfo {author} {\bibfnamefont
  {R.}~\bibnamefont {Gross}}, \ and\ \bibinfo {author} {\bibfnamefont
  {S.~T.~B.}\ \bibnamefont {Goennenwein}},\ }\href {\doibase
  10.1103/PhysRevLett.111.127003} {\bibfield  {journal} {\bibinfo  {journal}
  {Phys. Rev. Lett.}\ }\textbf {\bibinfo {volume} {111}},\ \bibinfo {pages}
  {127003} (\bibinfo {year} {2013})}\BibitemShut {NoStop}%
\bibitem [{\citenamefont {Kubo}\ \emph {et~al.}(2011)\citenamefont {Kubo},
  \citenamefont {Grezes}, \citenamefont {Dewes}, \citenamefont {Umeda},
  \citenamefont {Isoya}, \citenamefont {Sumiya}, \citenamefont {Morishita},
  \citenamefont {Abe}, \citenamefont {Onoda}, \citenamefont {Ohshima},
  \citenamefont {Jacques}, \citenamefont {Dr\'eau}, \citenamefont {Roch},
  \citenamefont {Diniz}, \citenamefont {Auffeves}, \citenamefont {Vion},
  \citenamefont {Esteve},\ and\ \citenamefont {Bertet}}]{cite07}%
  \BibitemOpen
  \bibfield  {author} {\bibinfo {author} {\bibfnamefont {Y.}~\bibnamefont
  {Kubo}}, \bibinfo {author} {\bibfnamefont {C.}~\bibnamefont {Grezes}},
  \bibinfo {author} {\bibfnamefont {A.}~\bibnamefont {Dewes}}, \bibinfo
  {author} {\bibfnamefont {T.}~\bibnamefont {Umeda}}, \bibinfo {author}
  {\bibfnamefont {J.}~\bibnamefont {Isoya}}, \bibinfo {author} {\bibfnamefont
  {H.}~\bibnamefont {Sumiya}}, \bibinfo {author} {\bibfnamefont
  {N.}~\bibnamefont {Morishita}}, \bibinfo {author} {\bibfnamefont
  {H.}~\bibnamefont {Abe}}, \bibinfo {author} {\bibfnamefont {S.}~\bibnamefont
  {Onoda}}, \bibinfo {author} {\bibfnamefont {T.}~\bibnamefont {Ohshima}},
  \bibinfo {author} {\bibfnamefont {V.}~\bibnamefont {Jacques}}, \bibinfo
  {author} {\bibfnamefont {A.}~\bibnamefont {Dr\'eau}}, \bibinfo {author}
  {\bibfnamefont {J.-F.}\ \bibnamefont {Roch}}, \bibinfo {author}
  {\bibfnamefont {I.}~\bibnamefont {Diniz}}, \bibinfo {author} {\bibfnamefont
  {A.}~\bibnamefont {Auffeves}}, \bibinfo {author} {\bibfnamefont
  {D.}~\bibnamefont {Vion}}, \bibinfo {author} {\bibfnamefont {D.}~\bibnamefont
  {Esteve}}, \ and\ \bibinfo {author} {\bibfnamefont {P.}~\bibnamefont
  {Bertet}},\ }\href {\doibase 10.1103/PhysRevLett.107.220501} {\bibfield
  {journal} {\bibinfo  {journal} {Phys. Rev. Lett.}\ }\textbf {\bibinfo
  {volume} {107}},\ \bibinfo {pages} {220501} (\bibinfo {year}
  {2011})}\BibitemShut {NoStop}%
\bibitem [{\citenamefont {Wallquist}\ \emph {et~al.}(2009)\citenamefont
  {Wallquist}, \citenamefont {Hammerer}, \citenamefont {Rabl}, \citenamefont
  {Lukin},\ and\ \citenamefont {Zoller}}]{cite11}%
  \BibitemOpen
  \bibfield  {author} {\bibinfo {author} {\bibfnamefont {M.}~\bibnamefont
  {Wallquist}}, \bibinfo {author} {\bibfnamefont {K.}~\bibnamefont {Hammerer}},
  \bibinfo {author} {\bibfnamefont {P.}~\bibnamefont {Rabl}}, \bibinfo {author}
  {\bibfnamefont {M.}~\bibnamefont {Lukin}}, \ and\ \bibinfo {author}
  {\bibfnamefont {P.}~\bibnamefont {Zoller}},\ }\href
  {http://stacks.iop.org/1402-4896/2009/i=T137/a=014001} {\bibfield  {journal}
  {\bibinfo  {journal} {Physica Scripta}\ }\textbf {\bibinfo {volume} {2009}},\
  \bibinfo {pages} {014001} (\bibinfo {year} {2009})}\BibitemShut {NoStop}%
\bibitem [{\citenamefont {Xiang}\ \emph {et~al.}(2013)\citenamefont {Xiang},
  \citenamefont {L\"u}, \citenamefont {Li}, \citenamefont {You},\ and\
  \citenamefont {Nori}}]{cite12}%
  \BibitemOpen
  \bibfield  {author} {\bibinfo {author} {\bibfnamefont {Z.-L.}\ \bibnamefont
  {Xiang}}, \bibinfo {author} {\bibfnamefont {X.-Y.}\ \bibnamefont {L\"u}},
  \bibinfo {author} {\bibfnamefont {T.-F.}\ \bibnamefont {Li}}, \bibinfo
  {author} {\bibfnamefont {J.~Q.}\ \bibnamefont {You}}, \ and\ \bibinfo
  {author} {\bibfnamefont {F.}~\bibnamefont {Nori}},\ }\href {\doibase
  10.1103/PhysRevB.87.144516} {\bibfield  {journal} {\bibinfo  {journal} {Phys.
  Rev. B}\ }\textbf {\bibinfo {volume} {87}},\ \bibinfo {pages} {144516}
  (\bibinfo {year} {2013})}\BibitemShut {NoStop}%
\bibitem [{\citenamefont {Verd\'u}\ \emph {et~al.}(2009)\citenamefont
  {Verd\'u}, \citenamefont {Zoubi}, \citenamefont {Koller}, \citenamefont
  {Majer}, \citenamefont {Ritsch},\ and\ \citenamefont
  {Schmiedmayer}}]{cite10}%
  \BibitemOpen
  \bibfield  {author} {\bibinfo {author} {\bibfnamefont {J.}~\bibnamefont
  {Verd\'u}}, \bibinfo {author} {\bibfnamefont {H.}~\bibnamefont {Zoubi}},
  \bibinfo {author} {\bibfnamefont {C.}~\bibnamefont {Koller}}, \bibinfo
  {author} {\bibfnamefont {J.}~\bibnamefont {Majer}}, \bibinfo {author}
  {\bibfnamefont {H.}~\bibnamefont {Ritsch}}, \ and\ \bibinfo {author}
  {\bibfnamefont {J.}~\bibnamefont {Schmiedmayer}},\ }\href {\doibase
  10.1103/PhysRevLett.103.043603} {\bibfield  {journal} {\bibinfo  {journal}
  {Phys. Rev. Lett.}\ }\textbf {\bibinfo {volume} {103}},\ \bibinfo {pages}
  {043603} (\bibinfo {year} {2009})}\BibitemShut {NoStop}%
\bibitem [{\citenamefont {Tyryshkin}\ \emph {et~al.}(2012)\citenamefont
  {Tyryshkin}, \citenamefont {Tojo}, \citenamefont {Morton}, \citenamefont
  {Riemann}, \citenamefont {Abrosimov}, \citenamefont {Becker}, \citenamefont
  {Pohl}, \citenamefont {Schenkel}, \citenamefont {Thewalt}, \citenamefont
  {Itoh},\ and\ \citenamefont {Lyon}}]{tyryshkin2012}%
  \BibitemOpen
  \bibfield  {author} {\bibinfo {author} {\bibfnamefont {A.~M.}\ \bibnamefont
  {Tyryshkin}}, \bibinfo {author} {\bibfnamefont {S.}~\bibnamefont {Tojo}},
  \bibinfo {author} {\bibfnamefont {J.~J.~L.}\ \bibnamefont {Morton}}, \bibinfo
  {author} {\bibfnamefont {H.}~\bibnamefont {Riemann}}, \bibinfo {author}
  {\bibfnamefont {N.~V.}\ \bibnamefont {Abrosimov}}, \bibinfo {author}
  {\bibfnamefont {P.}~\bibnamefont {Becker}}, \bibinfo {author} {\bibfnamefont
  {H.-J.}\ \bibnamefont {Pohl}}, \bibinfo {author} {\bibfnamefont
  {T.}~\bibnamefont {Schenkel}}, \bibinfo {author} {\bibfnamefont {M.~L.~W.}\
  \bibnamefont {Thewalt}}, \bibinfo {author} {\bibfnamefont {K.~M.}\
  \bibnamefont {Itoh}}, \ and\ \bibinfo {author} {\bibfnamefont {S.~A.}\
  \bibnamefont {Lyon}},\ }\bibfield  {title} {\enquote {\bibinfo {title}
  {Electron spin coherence exceeding seconds in high-purity silicon},}\ }\href
  {http://dx.doi.org/10.1038/nmat3182} {\bibfield  {journal} {\bibinfo
  {journal} {Nat Mater}\ }\textbf {\bibinfo {volume} {11}},\ \bibinfo {pages}
  {143--147} (\bibinfo {year} {2012})}\BibitemShut {NoStop}%
\bibitem [{\citenamefont {DiVincenzo}\ \emph {et~al.}(2000)\citenamefont
  {DiVincenzo}, \citenamefont {Bacon}, \citenamefont {Kempe}, \citenamefont
  {Burkard},\ and\ \citenamefont {Whaley}}]{cite03}%
  \BibitemOpen
  \bibfield  {author} {\bibinfo {author} {\bibfnamefont {D.~P.}\ \bibnamefont
  {DiVincenzo}}, \bibinfo {author} {\bibfnamefont {D.}~\bibnamefont {Bacon}},
  \bibinfo {author} {\bibfnamefont {J.}~\bibnamefont {Kempe}}, \bibinfo
  {author} {\bibfnamefont {G.}~\bibnamefont {Burkard}}, \ and\ \bibinfo
  {author} {\bibfnamefont {K.~B.}\ \bibnamefont {Whaley}},\ }\href
  {http://www.nature.com/nature/journal/v408/n6810/abs/408339a0.html}
  {\bibfield  {journal} {\bibinfo  {journal} {Nature}\ }\textbf {\bibinfo
  {volume} {408}},\ \bibinfo {pages} {339--342} (\bibinfo {year}
  {2000})}\BibitemShut {NoStop}%
\bibitem [{\citenamefont {Wolfowicz}\ \emph {et~al.}(2013)\citenamefont
  {Wolfowicz}, \citenamefont {Tyryshkin}, \citenamefont {George}, \citenamefont
  {Riemann}, \citenamefont {Abrosimov}, \citenamefont {Becker}, \citenamefont
  {Pohl}, \citenamefont {Thewalt}, \citenamefont {Lyon},\ and\ \citenamefont
  {Morton}}]{wolfowicz2013}%
  \BibitemOpen
  \bibfield  {author} {\bibinfo {author} {\bibfnamefont {G.}~\bibnamefont
  {Wolfowicz}}, \bibinfo {author} {\bibfnamefont {A.~M.}\ \bibnamefont
  {Tyryshkin}}, \bibinfo {author} {\bibfnamefont {R.~E.}\ \bibnamefont
  {George}}, \bibinfo {author} {\bibfnamefont {H.}~\bibnamefont {Riemann}},
  \bibinfo {author} {\bibfnamefont {N.~V.}\ \bibnamefont {Abrosimov}}, \bibinfo
  {author} {\bibfnamefont {P.}~\bibnamefont {Becker}}, \bibinfo {author}
  {\bibfnamefont {H.-J.}\ \bibnamefont {Pohl}}, \bibinfo {author}
  {\bibfnamefont {M.~L.~W.}\ \bibnamefont {Thewalt}}, \bibinfo {author}
  {\bibfnamefont {S.~A.}\ \bibnamefont {Lyon}}, \ and\ \bibinfo {author}
  {\bibfnamefont {J.~J.~L.}\ \bibnamefont {Morton}},\ }\bibfield  {title}
  {\enquote {\bibinfo {title} {Atomic clock transitions in silicon-based spin
  qubits},}\ }\href {http://dx.doi.org/10.1038/nnano.2013.117} {\bibfield
  {journal} {\bibinfo  {journal} {Nat Nano}\ }\textbf {\bibinfo {volume} {8}},\
  \bibinfo {pages} {561--564} (\bibinfo {year} {2013})}\BibitemShut {NoStop}%
\bibitem [{\citenamefont {George}\ \emph {et~al.}(2010)\citenamefont {George},
  \citenamefont {Witzel}, \citenamefont {Riemann}, \citenamefont {Abrosimov},
  \citenamefont {N\"otzel}, \citenamefont {Thewalt},\ and\ \citenamefont
  {Morton}}]{george2010}%
  \BibitemOpen
  \bibfield  {author} {\bibinfo {author} {\bibfnamefont {R.~E.}\ \bibnamefont
  {George}}, \bibinfo {author} {\bibfnamefont {W.}~\bibnamefont {Witzel}},
  \bibinfo {author} {\bibfnamefont {H.}~\bibnamefont {Riemann}}, \bibinfo
  {author} {\bibfnamefont {N.~V.}\ \bibnamefont {Abrosimov}}, \bibinfo {author}
  {\bibfnamefont {N.}~\bibnamefont {N\"otzel}}, \bibinfo {author}
  {\bibfnamefont {M.~L.~W.}\ \bibnamefont {Thewalt}}, \ and\ \bibinfo {author}
  {\bibfnamefont {J.~J.~L.}\ \bibnamefont {Morton}},\ }\bibfield  {title}
  {\enquote {\bibinfo {title} {Electron spin coherence and electron nuclear
  double resonance of bi donors in natural si},}\ }\href {\doibase
  10.1103/PhysRevLett.105.067601} {\bibfield  {journal} {\bibinfo  {journal}
  {Phys. Rev. Lett.}\ }\textbf {\bibinfo {volume} {105}},\ \bibinfo {pages}
  {067601} (\bibinfo {year} {2010})}\BibitemShut {NoStop}%
\bibitem [{\citenamefont {Benningshof}\ \emph {et~al.}(2013)\citenamefont
  {Benningshof}, \citenamefont {Mohebbi}, \citenamefont {Taminiau},
  \citenamefont {Miao},\ and\ \citenamefont {Cory}}]{cite01}%
  \BibitemOpen
  \bibfield  {author} {\bibinfo {author} {\bibfnamefont {O.}~\bibnamefont
  {Benningshof}}, \bibinfo {author} {\bibfnamefont {H.}~\bibnamefont
  {Mohebbi}}, \bibinfo {author} {\bibfnamefont {I.}~\bibnamefont {Taminiau}},
  \bibinfo {author} {\bibfnamefont {G.}~\bibnamefont {Miao}}, \ and\ \bibinfo
  {author} {\bibfnamefont {D.}~\bibnamefont {Cory}},\ }\href {\doibase
  http://dx.doi.org/10.1016/j.jmr.2013.01.010} {\bibfield  {journal} {\bibinfo
  {journal} {Journal of Magnetic Resonance}\ }\textbf {\bibinfo {volume}
  {230}},\ \bibinfo {pages} {84 -- 87} (\bibinfo {year} {2013})}\BibitemShut
  {NoStop}%
\bibitem [{\citenamefont {De~Graaf}\ and\ \citenamefont
  {Nicolay}(1997)}]{cite02}%
  \BibitemOpen
  \bibfield  {author} {\bibinfo {author} {\bibfnamefont {R.~A.}\ \bibnamefont
  {De~Graaf}}\ and\ \bibinfo {author} {\bibfnamefont {K.}~\bibnamefont
  {Nicolay}},\ }\href {\doibase
  10.1002/(SICI)1099-0534(1997)9:4<247::AID-CMR4>3.0.CO;2-Z} {\bibfield
  {journal} {\bibinfo  {journal} {Concepts in Magnetic Resonance}\ }\textbf
  {\bibinfo {volume} {9}},\ \bibinfo {pages} {247--268} (\bibinfo {year}
  {1997})}\BibitemShut {NoStop}%
\bibitem [{\citenamefont {Garwood}\ and\ \citenamefont {Ke}(1991)}]{cite04}%
  \BibitemOpen
  \bibfield  {author} {\bibinfo {author} {\bibfnamefont {M.}~\bibnamefont
  {Garwood}}\ and\ \bibinfo {author} {\bibfnamefont {Y.}~\bibnamefont {Ke}},\
  }\href {\doibase http://dx.doi.org/10.1016/0022-2364(91)90137-I} {\bibfield
  {journal} {\bibinfo  {journal} {Journal of Magnetic Resonance (1969)}\
  }\textbf {\bibinfo {volume} {94}},\ \bibinfo {pages} {511 -- 525} (\bibinfo
  {year} {1991})}\BibitemShut {NoStop}%
\bibitem [{\citenamefont {Kupce}\ and\ \citenamefont {Freeman}(1995)}]{cite08}%
  \BibitemOpen
  \bibfield  {author} {\bibinfo {author} {\bibfnamefont {E.}~\bibnamefont
  {Kupce}}\ and\ \bibinfo {author} {\bibfnamefont {R.}~\bibnamefont
  {Freeman}},\ }\href {\doibase http://dx.doi.org/10.1006/jmra.1995.1179}
  {\bibfield  {journal} {\bibinfo  {journal} {Journal of Magnetic Resonance,
  Series A}\ }\textbf {\bibinfo {volume} {115}},\ \bibinfo {pages} {273 -- 276}
  (\bibinfo {year} {1995})}\BibitemShut {NoStop}%
\bibitem [{\citenamefont {Becker}\ \emph {et~al.}(2010)\citenamefont {Becker},
  \citenamefont {Pohl}, \citenamefont {Riemann},\ and\ \citenamefont
  {Abrosimov}}]{cite14}%
  \BibitemOpen
  \bibfield  {author} {\bibinfo {author} {\bibfnamefont {P.}~\bibnamefont
  {Becker}}, \bibinfo {author} {\bibfnamefont {H.-J.}\ \bibnamefont {Pohl}},
  \bibinfo {author} {\bibfnamefont {H.}~\bibnamefont {Riemann}}, \ and\
  \bibinfo {author} {\bibfnamefont {N.}~\bibnamefont {Abrosimov}},\ }\href
  {\doibase 10.1002/pssa.200925148} {\bibfield  {journal} {\bibinfo  {journal}
  {physica status solidi (a)}\ }\textbf {\bibinfo {volume} {207}},\ \bibinfo
  {pages} {49--66} (\bibinfo {year} {2010})}\BibitemShut {NoStop}%
\bibitem [{\citenamefont {Blank}\ \emph {et~al.}(2013)\citenamefont {Blank},
  \citenamefont {Dikarov}, \citenamefont {Shklyar},\ and\ \citenamefont
  {Twig}}]{cite13}%
  \BibitemOpen
  \bibfield  {author} {\bibinfo {author} {\bibfnamefont {A.}~\bibnamefont
  {Blank}}, \bibinfo {author} {\bibfnamefont {E.}~\bibnamefont {Dikarov}},
  \bibinfo {author} {\bibfnamefont {R.}~\bibnamefont {Shklyar}}, \ and\
  \bibinfo {author} {\bibfnamefont {Y.}~\bibnamefont {Twig}},\ }\href {\doibase
  http://dx.doi.org/10.1016/j.physleta.2013.05.025} {\bibfield  {journal}
  {\bibinfo  {journal} {Physics Letters A}\ }\textbf {\bibinfo {volume}
  {377}},\ \bibinfo {pages} {1937 -- 1942} (\bibinfo {year}
  {2013})}\BibitemShut {NoStop}%
\end{thebibliography}%

\end{document}